# Data Deluge in Astrophysics: Photometric Redshifts as a Template Use Case

Massimo Brescia [1][0000−0001−9506−5680], Stefano Cavuoti [1,2,3][0000−0002−3787−4196],

Valeria Amaro [2][0000−0003−1953−5104], Giuseppe Riccio [1][0000−0001−7020−1172],

Giuseppe Angora [1,2][0000−0002−0316−6562], Civita Vellucci [2][0000−0002−8141−5552],

and Giuseppe Longo [2,3][0000−0002−9182−8414]

[1] INAF - Osservatorio Astronomico di Capodimonte, via Moiariello 16, I-80131,

Napoli, Italia,

[2] Università degli Studi Federico II - Dipartimento di Fisica "E. Pancini", via Cintia 6,

I-80135 Napoli, Italia

[3] INFN - Napoli Unit, via Cintia 6, I-80135 Napoli, Italia

brescia@oacn.inaf.it

**Abstract.** Astronomy has entered the big data era and Machine Learning based methods have found widespread use in a large variety of astronomical applications. This is demonstrated by the recent huge increase in the number of publications making use of this new approach. The usage of machine learning methods, however is still far from trivial and many problems still need to be solved. Using the evaluation of photometric redshifts as a case study, we outline the main problems and some ongoing efforts to solve them.

## 1. The Astronomical Data Deluge

Astronomy has entered the big data era with ground-based and space-borne observing facilities producing data sets in the Tera-byte and Peta-byte domain and, in the near future, instruments such as LSST (Large Synoptic Survey Telescope [1]) and SKA (Square Kilometer Array [2]), to quote just a few, will produce data streams of unprecedented size and complexity [3]. The need to store, reduce, analyze and to extract useful information from this unprecedented deluge of data has triggered the growth of the new discipline of Astroinformatics, placed at the intersection among Information and Communication Technology, Statistics, and Astrophysics, which aims at provid-ing the community with the needed infrastructures and know how. The advent of Astroinformatics is currently driving a true revolution in the methodology of astronomical research. Wordings like "data mining", "machine learning" (or ML), "Bayesian statistics", which rarely appeared in the literature of the last century, have suddenly become very popular in the recent astronomical literature. A trend which will be shortly exemplified in the next section (Sec. 2) where we shall also show the frequency with which these words have been used in the recent literature. In the remaining of this paper we shall instead discuss some problems that are commonly encountered in the application of ML methods to astronomical problems, using the evaluation of photometric redshifts as a case study.

This article is the full version of our work, and contains well over the 30% of the extended abstract presented at the DAMDID 2017 Conference and published in the related Proceedings [4].

## 2. The fast uptake of Astroinformatics

In almost all fields, modern technology allows to capture and store huge quantities of heterogeneous and complex data often consisting of hundreds of features for each record and requiring complex metadata structures to understand. This has led to a situation well described by the following sentence: *...while data doubles every year, useful information seems to be decreasing, creating a growing gap between the generation of data and our understanding of it...* [5]. A need has therefore emerged for a new generation of analytics tools, largely automatic, scalable and highly reliable. Strictly speaking, Knowledge Discovery in Databases (KDD) is about algorithms for inferring knowledge from data and ways of validating the obtained results [6], as well as about running them on infrastructures able to match the computational demands. In practice, whenever there is too much data or, more generally, a representation in more than 4 dimensions [7], there are basically three ways to make learning feasible. The first one is straightforward: applying the training scheme to a decimated dataset.

The second method relies on parallelization techniques, the idea being to split the problem into smaller parts, then solving each one by using a separate CPU and finally combining the results together [8]. Sometimes this is feasible due to the intrinsic natural essence of the learning rule (such as genetic algorithms; [9]). However, even after parallelization, the algorithm's asymptotic time complexity cannot be improved. The third and more challenging way to enable a learning paradigm to deal with massive data sets is to develop new algorithms of lower computational complexity, but in many case this is simply unfeasible [10]. Such complexity is one of the main explanations for the gap still existing between the new methodologies and the wide community of potential users who fail to adopt them. In order to be effective, in fact, the KDD methodology requires a good understanding of the mathematics and statistics underlying the methods and learning paradigms, of the computing infrastructures and of the complex workflows that need to be implemented. Most users (even in the scientific community) are usually not willing to make the effort to understand the process and prefer to recur to traditional approaches that are far less powerful, but much more affordable [11]. Astroinformatics copes with these concepts, by exploiting KDD methodologies and by taking into account the fact that background knowledge can make possible to reduce the amount of data to process. It adopts data mining and machine learning paradigms, mostly based on the assumption that in many cases a bunch of the parameter space attributes turn out to be irrelevant when background knowledge is taken into account [8]. Dimensional reduction, classification, highly non-linear regression, prediction, clustering, filtering, are all concrete examples of functionalities belonging to the KDD conceptual domain, in which the various Astroinformatics methodologies can be applied to explore data under peculiar aspects,

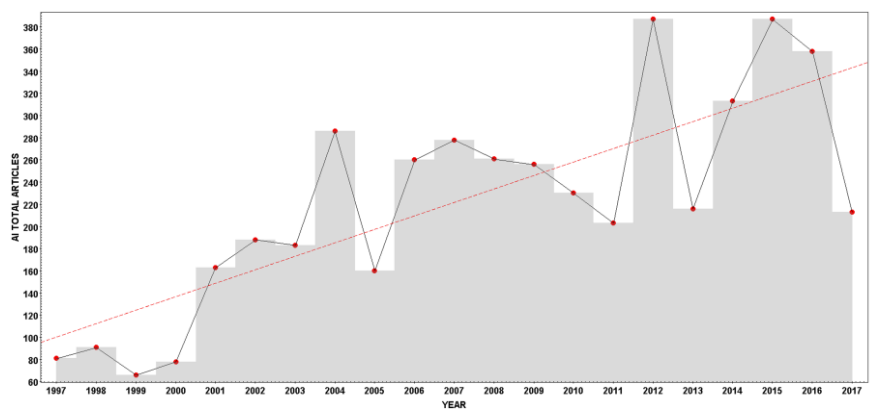

**Fig.1.** Quantity of scientific articles focusing Astroinformatics topics, published per year on specialized refereed journals, from January 1997 to July 2017. The dotted red line is the linear regression trend.

strictly connected to the associated functionality scope [12].

In the current Big Data analytics scenario, astronomical observations and related data gathering, reduction and analysis represent one of the clearest examples of *data-driven* science, where the amount of data collected in a single day is enough to keep occupied an entire community of scientists for the rest of their life. Therefore, in the last twenty years an awareness has been reached and consolidated in the scientific community: that the union can make the difference, heterogeneity is not dispersion, multi-disciplinary eclecticism is not a presumptuous ambition, but rather they are unavoidable and precious qualities. A characterizing factor of the astrophysical data is that behind them there are specific physical laws supervising complex phenomena. Machine learning methods, by achieving a higher level of complexity, will surely allow disclosing patterns and trends pointing at a higher level of complexity than what has been achieved so far. In fact, in the last decade, in many different fields it has been clearly demonstrated that the emulation of the mechanisms underlying the intelligence in Nature (an expression of the human brain functioning, the sensory organs and the formulation of thoughts), if translated into efficient algorithms and supplied to super computers, is fully and rapidly able to analyze, correlate and extract huge amounts of heterogeneous information. Astroinformatics has grown in the last twenty years precisely as a science identifying the symbiosis between different scientific and technological disciplines, therefore a perfect example of a virtuous resource, through the combination of Astronomy, Astrophysics, Artificial Intelligence, Data Mining and Computer Science. The growing awareness of the need to do science without being submerged by the complexity and quantity of the information to be analyzed is amply demonstrated by the result of our recent bibliographic survey (Fig. 1), in which the positive tendency to increase the study and applicability of these methodologies in the astrophysical field is more than evident. We are hence firmly convinced that the future of Astrophysics, like many other scientific and social disciplines, cannot ignore the awareness that the stereotype of the scientist in the Galilean conception, valid up to twenty years ago, is now undergoing an evolutionary jump. The contemporary scientist must necessarily assume an "open minded" role, must be diversified in professional qualifications, master of all the communicative mechanisms available in the digital age, and with a solid openness and eclectic ability of Darwinian adaptation to the new demands of modern scientific exploitation.

## 3. A template case: Photometric redshifts

Photometric redshifts (photo-z), by allowing the calculation of the distances for large samples of galaxies, are at the very heart of modern cosmological surveys. The main idea behind photo-z is that there exists a highly nonlinear correlation between broad-band photometry and redshift; a correlation arising from the fact that the stretching introduced by the cosmological redshift causes the main features in a galaxy spectrum to shift through the different filters of a given photometric system [13,14]. In a rough oversimplification, there are two classes of methods to derive photo-z: the Spectral Energy Distribution (SED) template fitting methods (e.g. [15–18]) and empirical (or interpolative) methods (e.g. [19–23]).

SED methods rely on fitting the multi-wavelength photometric observations of the objects to a library of synthetic or observed template SEDs, which are first convolved with the transmission functions of the given filters and then shifted to create synthetic magnitudes for each galaxy template as a function of the redshift. SED methods are capable to derive photo-z, the spectral type and the Probability Density Function (hereafter PDF) of the photo-z error distribution for each source, all at once. However, they suffer from several shortcomings, such as in particular, the potential mismatch between the templates used for the fitting, and the properties of the selected sample of galaxies [24], color/redshift degeneracies, and template incompleteness.

Among empirical methods, those based on various Machine Learning (ML) algorithms are more commonly used to infer the hidden relation between input (mostly multi-band photometry) and the desired output (redshift). In supervised ML techniques, the learning process is regulated by the spectroscopic redshift available for a subsample of the objects, forming what is known as the Knowledge Base or KB. In the unsupervised approach, the spectroscopic information is not used in the training phase, but only during the validation phase. Over the years many ML algorithms have been used to this purpose; to quote just a few: neural networks [19, 25], boosted decision trees [26], random forests [27], self organized maps [28], quasi Newton algorithm [29], genetic algorithms [30, 31]. Numerous examples (see for instance [32]) show that, when a suitable KB is available, most ML methods outperform the SED fitting ones. ML methods however, suffer from the difficulty of achieving good performances outside the regions of the observed Parameter Space (PS) properly covered by the KB. For instance, they are not effective in deriving the redshifts of objects fainter than the available spectroscopy. A few attempts have been made to overcome this limitation (cf. [33]), but with questionable success. Moreover, there are few examples of combined approaches involving both SED and ML methods (e.g. [34, 35]), which open promising perspectives. Since a detailed analysis of the vast literature is beyond the purpose of this work, we shall focus on some among the main issues still open. Namely: the problem of missing data; the definition of the optimal parameter space and the related problem of feature selection; the characterization of the parameter space in terms of completeness and accuracy of the results; and finally the derivation of reliable PDFs.

### 3.1 Missing data

Most ML methods are not robust against missing data (either missing observations or non-detections). In such cases there are two main solution strategies: list-wise deletion and imputation. The list-wise deletion consists basically in the elimination of all the entries with one (at least) missing value; clearly this is the simplest and most commonly used approach [20–22, 36, 37]. However, it has the main drawback of reducing the number of objects for which it is possible to predict the output and this is quite problematic in astronomical applications, especially for objects at high redshift, where a large fraction of objects is not detected in at least one band.

Imputation, instead, consists in replacing the missing data (of any kind) with substituted value. Obviously, the label imputation is just an umbrella under which there exists a plethora of methods. The simplest type of imputation is the usage of the same value for all the entries with a missing data (i.e. −99) that actually creates portions of the parameter space (one for each feature with at least one missing value), in which all such objects will fall. A slightly different approach consists in the substitution of a realistic value. We consider as realistic a value that belongs to the domain of the feature itself, i.e. its minimum, maximum, mean or median. This method can be effective either when a parameter has not been measured for some reason or when, rather than with missing data, the user is dealing with non-detections (e.g. upper limits in the flux). Another approach consists in a sort of hierarchical regression on all features: in practice for each pattern with at least one missing value, the latter is replaced with a prediction based on the other elements of the dataset. Such replacement can be operated either with traditional interpolation or, more effectively, with any ML methods (such as kNN; Cavuoti et al., in preparation). To the best of our knowledge, only a few attempts in this direction have been made on astronomical data sets in spite of the fact that an effective imputation would greatly increase the number of objects for which photo-z could be derived (even though with lower accuracy).

### 3.2 Biases in the Parameter Space

The performance of any supervised method ultimately depend on how well the training set maps the observed parameter space. From a qualitative point of view, this implies that any ML method will fail or have very low performances either when applied to objects of a missing type in the training set, or when applied to objects falling in regions of the PS under-sampled in the training set.

Focusing on the latter problem, an interesting approach, to robustly map the empirical distribution of galaxies in a multidimensional colour space, has been presented in [38] under the form of unsupervised clustering (via Self Organizing Maps) of the spectroscopic objects in the COSMOS survey [39]. In order to prepare the ground for the future Euclid surveys [40], the photometry was reduced to the Euclid photometric system, and all galaxies in the COSMOS master catalogue were clustered (in a completely data driven approach) to find similar objects. In this way, they found that (cf. Fig. 6 in [38]) more than half of the parameter space was poorly or not covered by training data. Additionally, they found also the first evidence that the colour-redshift relation may be smoother than what is commonly believed on the basis of the photo-z variance estimates derived from template fitting. If confirmed, this result would prove that ML photo-z are intrinsically more robust than those derived with SED fitting techniques. High intrinsic variance in the colour-redshift mapping should in fact have resulted in a large cell-to-cell variation in median spec-z, whereas the actual distribution appeared to be rather smooth overall.

### 3.3 Optimal feature selection

Feature selection, i.e. the choice of the optimal number and combination of parameters to be used in the training of a ML based method, is a crucial step often overlooked on the basis of the false belief that the more are the parameters, the better are the results. This is well known in the data mining literature as "the curse of dimensionality" that can be shortly summarized as it follows: for a fixed number of data points in the training set, the density of training points in a given voxel of the parameter space decreases for increasing number of parameters. Beyond a given threshold, this leads to a decrease in information content and an increase in noise that in turn lowers the performances. Unfortunately, there is not a unique way to find this optimal combination of parameters.

A "Data Driven" approach has been recently explored in [37], by using a data set extracted from the Sloan Digital Sky Survey data release 7. They run two different experiments: in the first one, they explored all possible combinations of four parameters extracted from a subset of 55 parameters provided by the Sloan Digital Sky Survey (SDSS, [41]) archive in order to see whether it was possible to improve the results obtained in [36]. They found that features, selected without relying on the opinion of an expert, improved the results by almost 30%. In a second experiment [42] they used 90 parameters (magnitudes, errors, radii and ellipticities) from the SDSS database and combine them in unusual ways (average, difference, ratio etc.) obtaining a set of ∼ 5000 features. Then they run a forward selection looking for the features that optimized the root mean square error (RMSE). The data driven approach, however, is very demanding in terms of computational time and for most real life applications other approaches need to be explored.

The most surprising result was that, at least at first glance, in both experiments the optimal combination of features makes no sense to the average astronomer. In other words, they proved that the features containing the most information would have never been selected by a human expert.

The coherence between any optimal parameter space found by feature selection and its physical legitimacy is an aspect still poorly explored in the astronomical literature. But for sure it is something to keep in mind while implementing the next generation pipelines for data reduction and analysis.

### 3.4 The evaluation of Probability Density Functions

In the context of photo-z and in absence of systematics, the factors affecting the final reliability of the estimates obtained with ML methods are photometric errors, internal errors of the methods and statistical biases in the photometric parameter space or in the KB. All these factors can be summarized by deriving a PDF for both the individual measurements and for large samples of objects (cumulative PDF). We wish to underline that, from a rigorous statistical point of view, a PDF should be an intrinsic property of a certain phenomenon, regardless the adopted method used to quantify the phenomenon itself. However, in the case of photo-z, the PDF turns out to be strictly dependent both on the measurement methods and on the underlying physical assumptions. Such simple statistical consideration raises some doubts on the true meaning of PDFs derived with any photo-z method.

The derivation of PDFs with machine learning method has been subject of many attempts, (cf. [43]) but in most cases, the PDF took into account only the internal errors and biases of the methods (arising from the random initialization of the weights and/or the random extraction of the training sample from the KB).

A method capable to derive a reliable PDF, which takes into account all sources of errors, is METAPHOR (Machine learning Estimation Tool for Accurate PHOtometric Redshifts; [44]).

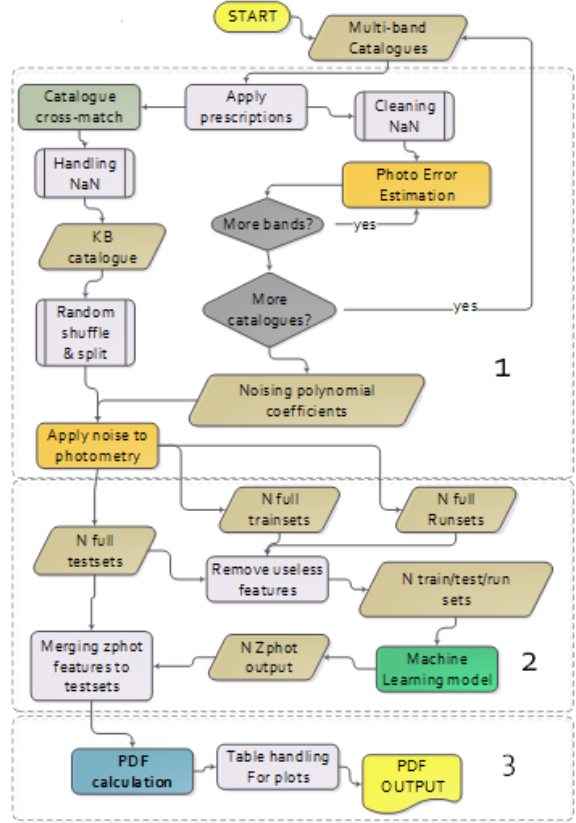

**Fig.2.** The processing flow of the METAPHOR pipeline to estimate photometric redshifts by means of machine learning methods and to provide photo-z PDFs.

It was designed as a modular workflow capable to plug-in different ML models (Fig. 2). In order to estimate internal errors, this method runs N trainings on the same training set, or M trainings on M different random extractions from the KB, to derive the initialization errors and the biases induced by the randomness of the training set, respectively. The main advantage of METAPHOR is the possibility to take into account the photometric flux errors: as extensively explained in [44], this is achieved by introducing a proper perturbation (either measured or parametrized from the error distribution) on the photometric data for the objects in test set. METAPHOR has been successfully tested on SDSS-DR9 [44] and the ESO Kilo-Degree Survey Data Release 3 (KiDS-ESO-DR3 [45]) using the MLPQNA model (Multi Layer Perceptron trained with Quasi Newton Algorithm; [10, 21, 46]) as the internal photo-z estimation ML engine. It is worth stressing, however, that the internal ML model can be easily replaced, thus allowing also METAPHOR as a tool to estimate the effects of different models on the final estimates.

In a recent work [47], three different sets of PDFs obtained with three different methods (METAPHOR; ANNz2 [43] and the SED fitting method BPZ [48]) for the same data set extracted from the KiDS-ESO DR3, were compared using several statistical indicators.

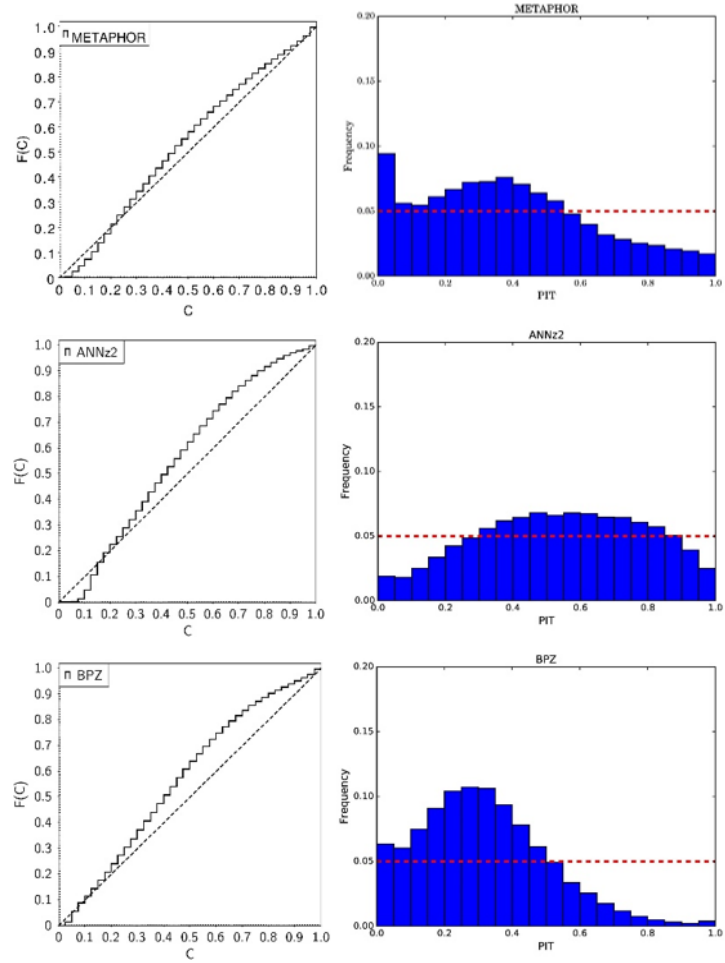

**Fig.3.** Wittmann Credibility Curve (left panels) and the PIT diagram (right panels) plots obtained on KiDS-ESO DR3 data for the PDFs provided by METAPHOR (upper panel), ANNz2 (middle), and BPZ (lower).

For instance, the PIT (Probability Integral Transform, [49]) histogram and the credibility analysis performed through the Wittmann [50] diagrams show that the three methods fail (even though in different ways) to capture the underlying redshift distribution (Fig. 3). PIT curves show in fact the total under-dispersive character of the reconstructed photometric redshifts distributions; while the Wittmann diagrams confirm an overconfidence of all photo-z estimates. Therefore, these results showed also that much work still remains to be done in order to fix a standard set of always reliable evaluation metrics.

## 4 Conclusions

Astroinformatics methods have found a widespread use in recent astrophysical contexts and a good knowledge of their methodological background is becoming essential for any observational astronomer. However, as it always happens, the adoption of a new methodology discloses new problems to solve. By using the derivation of photometric redshifts as template case, we pointed out some general issues that are and will be encountered in most, if not all, astroinformatics applications. However, it needs to be said that at least some of the problems currently encountered derive from the fact that the adoption of machine learning and KDD methodologies is still incomplete and at an early stage. For instance, the results in [37] clearly show that, for a given task, the parameters usually selected by astronomers are not necessarily those carrying most of useful information. This is not only a problem of proper feature selection, but it might require an entirely different and "data driven" approach to the definition and measurement of astronomical parameters. Deep learning can help in some applications but it is not the answer to all problems and it is very likely that in the near future it will become necessary to re-think and re-define the way to measure the parameters.

Furthermore, more approaches to the problems related to missing data need to be explored. For example, techniques based on the co-clustering (i.e. simultaneous clustering of the objects and of the features), but to the best of our knowledge, available methods are too much demanding in terms of computing time to be effectively used on very large data sets.

Finally, further investigation is required about the PDFs obtained with ML methods, in order to assess reproducible and reliable error estimates.

**Acknowledgements**
MB acknowledges the INAF PRIN-SKA 2017 program 1.05.01.88.04 and the funding from MIUR Premiale 2016: MITIC. MB and GL acknowledge the H2020-MSCA-ITN-2016 SUNDIAL (SUrvey Network for Deep Imaging Analysis and Learning), financed within the Call H2020-EU.1.3.1.